\documentclass[aps,prl,showpacs,english,10pt,a4,nofootinbib,notitlepage,twocolumn,superscriptaddress]{revtex4-1}

\usepackage[utf8]{inputenc}
\usepackage[T1]{fontenc}
\usepackage[english]{babel}

\usepackage{times}
\usepackage{dsfont}

\usepackage{amssymb,amsmath,amsfonts,amsthm}
\usepackage{mathtools}
\usepackage{verbatim}
\usepackage{enumerate}
\usepackage{graphicx}
\graphicspath{{figs/}}
\usepackage{hyperref}
\usepackage{bbm}
\usepackage{booktabs}

\usepackage[caption=false]{subfig}

\usepackage{color}
\definecolor{dred}{rgb}{.8,0.2,.2}
\definecolor{ddred}{rgb}{.8,0.5,.5}
\definecolor{dblue}{rgb}{.2,0.2,.8}
\definecolor{dgreen}{rgb}{.2,0.5,.2}

\newcommand{\bra}[1]{\mbox{$\langle #1|$}}
\newcommand{\ket}[1]{\ensuremath{|#1\rangle}}

\newcommand{\be}{\begin{equation}}
\newcommand{\ee}{\end{equation}}

\newcommand{\bea}{\begin{eqnarray}}
\newcommand{\eea}{\end{eqnarray}}

\begin{document}

\title{Preparation of Logically Labeled Pure States with Only Two Turns for Bulk Quantum Computation}
\date{\today}
\author{Tao Xin}
\affiliation{State Key Laboratory of Low-dimensional Quantum Physics and Department of Physics, Tsinghua University, Beijing 100084, China}
\affiliation{Tsinghua National Laboratory of Information Science and Technology,  Beijing 100084, China}

\author{Liang Hao}
\affiliation{Institute of Applied Physics and Computational Mathematics, Beijing, 100094, China}

\author{Shi-Yao Hou}
\affiliation{Microsystem and Terahertz Research Center, China Academy of Engineering Physics, Chengdu, 610200, China}
\affiliation{Institute of Electronic Engineering, China Academy of Engineering Physics, Mianyang, 621999, China}

\author{Guan-Ru Feng}
\affiliation{Institute for Quantum Computing, Waterloo, Ontario, N2L 3G1, Canada}
\affiliation{Department of Physics and Astronomy, University of Waterloo, Waterloo, Ontario, N2L 3G1, Canada}

\author{Gui-Lu Long}
\email[Correspondence and requests for materials should be addressed to G.L.L.: ]{gllong@tsinghua.edu.cn}
\affiliation{State Key Laboratory of Low-dimensional Quantum Physics and Department of Physics, Tsinghua University, Beijing 100084, China}
\affiliation{Tsinghua National Laboratory of Information Science and Technology,  Beijing 100084, China}
\affiliation{The Innovative Center of Quantum Matter, Beijing 100084, China}

\begin{abstract}
Quantum state preparation plays an equally important role with quantum operations and measurements in quantum information processing. The previous methods of preparing initial state for bulk quantum computation all have inevitable disadvantages, such as, requiring multiple experiments, causing loss of signals, or requiring molecules with restrictive structure. In this work, three kinds of quantum circuits are introduced to prepare the pseudo-pure states of ($n-1$) qubits in the Hilbert space of $n$ coupled spins  which merely need the assist of one ancilla spin and two experiments independent of $n$.  Being without gradient fields effectively avoids the reduction of the signals. Our methods have no special requirements on the structure of the used molecules. To test these methods more comprehensively,  we experimentally
demonstrate the preparation of the labeled pseudo-pure states using heteronuclear 2-qubit and homonuclear 4-qubit nuclear magnetic resonance quantum information processor.

\end{abstract}

\maketitle

\textit{Introduction. }-- 
Quantum computer, based on quantum mechanics, provides the extraordinary potential to solve certain problems in faster way which are usually intractable on classical physical computer \cite{simon,shor,grover,ekert,kubinec,Divincenzo,bennett,chuang}, which plays an important role in quantum simulations when dealing with the special problems, such as, dirac equation and quantum relativistic effects \cite{Gerritsma,Lamata}, quantum bakers map \cite{Weinstein}, chemical reactions \cite{daweic}, and molecular energies \cite{Aspuru,daweim}, as well as in quantum algorithm including the algorithm for finding eigenvalues and eigenvectors \cite{Abrams} and the algorithm for linear systems of equations \cite{Harrow}. The effective solution of these problems further leads to numerous ways to realize quantum computers, such as, trapped ions \cite{c9},
quantum dots \cite{c10}, cavity QED \cite{c11}, silicon-based nuclear spins \cite{c12}, superconducting Josephson junctions \cite{c13} and nuclear magnetic resonance (NMR) systems \cite{c14,chuang1}. As a bulk quantum computer, spins in NMR are undoubtedly well-established quantum computer processor and have readily available techniques compared with the other physical implementations. Further, techniques developed from NMR are as well available in other quantum systems  \cite{Gulde,Mintert}.

In bulk quantum computation, successful preparation of initial pure state, usually starting from a trivial high-mixed state which can not be the input state for quantum computing, is crucial for subsequent unitary operations and measurements  of expectation values of some observables. However, the concept of pseudo-pure state (PPS) is commonly used instead of true pure state \cite{c14}. A PPS has similar behaviors to a pure state when evolving under the Hamiltonians of NMR, which is supported by the fact that  the identity matrix is not observable in NMR spectroscopy and does not transform under any unitary operations. There have been a lot
of methods for preparing pseudo pure states over the past decade \cite{peng,Cory1, Knill,Vandersypen,Laflamme}. These approaches may be divided into spatial averaging \cite{Cory1}, temporal averaging \cite{Knill}, logical labeling \cite{Vandersypen}, and cat-state method \cite{Laflamme}. However, they always suffers from a number of practical disadvantages.

\begin{table}[tbp!]
\centering
\caption{\footnotesize{Comparison of the different methods for preparing the PPS.}} \label{comp}
\begin{tabular}{cccc}
\hline
\hline
Methods & Turns & Grad fields & Ancilla qubits \\
\hline
Spatial Averaging&1 	 &$f_s(n)$&0\\
Temporal Averaging & $f_t(n)$ 	 &0&0\\
Logical Labeling  &1	 &0&$f_l(n)$\\
Cat-State  &1	 &1+$f_c(n)$&1\\
LPPS-TT1(TT2)  &2	 &0&1\\
LPPS-TT3  &2	 &1&1\\
\hline
\hline
\end{tabular}
\end{table}

In this Letter, combining the properties and advantages of the temporal averaging \cite{Knill} and logical labeling approaches \cite{Vandersypen}, we subtly propose three kinds of methods for preparing a PPS, all of which require two experiments for a system with any number of dimensions and are applicable for both homo-nuclear and hetero-nuclear molecules. These methods can be used to prepare labeled pseudo-pure states based on two turns,  which are called LPPS-TT$m$ ($m=1,2,3$) for short in the following. LPPS-TT1 and LPPS-TT2 methods, without using gradient fields, can be used to prepare the PPS $(\ket{0}\bra{0}^{\otimes n}-\ket{1}\bra{1}^{\otimes n})$ and $(\ket{0}\bra{0}-\ket{1}\bra{1})\otimes \ket{0}\bra{0}^{\otimes (n-1)}$, respectively. These methods do not result in reduction of the signals and will be an inspiring option when high signal-noise ratio is needed in large system. LPPS-TT3 using one gradient field as well realizes the preparation of the PPS $(\ket{0}\bra{0}-\ket{1}\bra{1})\otimes \ket{0}\bra{0}^{\otimes (n-1)}$, of which the quantum circuit is simpler than that of LPPS-TT2. Table \ref{comp} fully demonstrates the comparison between the existing techniques and our methods for preparing PPS from different perspectives including the number of required experiments, gradient fields, and ancilla qubits, where $f(n)$ is an increasing function of $n$ and a subscript means the abbreviation of the method. To demonstrate the methods,  we implemented them on a heteronuclear 2-spin and a homonuclear 4-spin samples in room-temperature liquid NMR. Full state tomography is further implemented on the final state after LPPS-TT$m$ processing to evaluate the quality of the prepared PPS.

\textit{The algorithm. }-- 
In bulk quantum computation \cite{Chuang2,Chuang3}, such as liquid  NMR systems based on macroscopic ensembles of quantum spins, the preparation of
the desired input state for quantum computing is always from  a
thermal equilibrium state $\rho_{eq}=e^{-\mathcal{H}/k_{B}T}/Z$, with 
the Boltzmann constant $k_{B}$, the partition function
normalization factor $Z$ and the thermodynamic temperature $T$.  Under a strong
magnetic field $B_0$, the internal Hamiltonian $\mathcal{H}$ of $n$-spin system can be approximately written as ($\hbar=1$) \cite{Ernst},
\begin{equation}
\mathcal{H}_n=-\sum_{i}^{n} (\omega_i-2\pi\nu_i) I_z ^{i}+\sum_{i<j}^{n}2\pi J_{ij}I_z
^{i}I_z ^{j},
\end{equation}
where $I_z ^{i}=\sigma_z^{i}/2$ are the spin operators, $\omega_i=\gamma_i B_0$ and $\nu_i$ are respectively the Larmour frequency and the chemical shift of the $i$th spin with gyromagnetic ratio $\gamma_i$. $J_{ij}$ is the strength of the corresponding scalar spin-spin coupling
interaction. Considering that $|J_{ij}|<<\omega_i$ at room temperature, density matrix of the thermal equilibrium state can be approximated to
\begin{equation}\label{e1}
\rho_{eq}\approx\frac{1}{2^{n}}I^{\otimes n}+\frac{B_0}{2^{n}k_B
T}\sum_{i}^{n}\gamma_i I_z ^{i},
\end{equation}
where $I$ is the $2\times 2$ identity matrix. The thermal equilibrium state
is a highly mixed state, so it is impossible to perfectly align the spins at room temperature, which means we do not have the ability of preparing a true pure state in room temperature liquid NMR. Fortunately, we can introduce so-called pseudo pure state or effective pure state to avoid this problem  \cite{c14},  the density matrix of which is  
\begin{equation}\label{e2}
\rho_{eff}=\frac{1-\epsilon}{2^n}I^{\otimes n}+\epsilon \ket{0}\bra{0}^{\otimes n}.
\end{equation}
The reason that this new concept is convincing is that the big part $(1-\epsilon)I^{\otimes n}/2^n$ remains unchanged and that it does not contribute to NMR spectra under any unitary operations. Hence we only focus on the deviation density matrix $\Delta \rho=\ket{0}\bra{0}^{\otimes n}$ as the effective input state of quantum computing in liquid NMR. In the following, we present how to prepare such a PPS from a
thermal equilibrium state via the new methods LPPS-TT$m$ in detail.

In general,  we consider a system with $n$ 1/2-spins. The first spin of them is used as the ancilla qubit. The deviation density matrix of thermal equilibrium under the high temperature approximation can be described by $\rho^*_n=\sum_{i}^{n}\gamma_i I_z ^{i}$. The methods LPPS-TT$m$ for bulk quantum computation mainly include two steps. Step one, for LPPS-TT$m$ $(m=1,2)$, the thermal equilibrium state $\rho^*_n$ is directly used as the input state $\rho^m_n$ without additional operations. For LPPS-TT3 method, only the magnetization $\gamma_1 I^1_z$ of the ancilla spin is remained as  the input state $\rho^3_n$ which can be realized by applying a gradient field after a $\pi/2$ rotation on the work spins. Step two,  we attempt to reconstruct the unitary operation $U^m_{n}$ for redistributings the population of $\rho^m_n$, to obtain the result $\rho^m_{nu}=U^m_{n}\rho^*_nU^{m\dagger}_{n}$. Sum over the above density matrices, we will create the desired logically labeled PPS $\bar{\rho}^m_n=\rho^m_n+\rho^m_{nu}$.

\begin{figure}[htb]
\begin{center}
\includegraphics[width= 1\columnwidth]{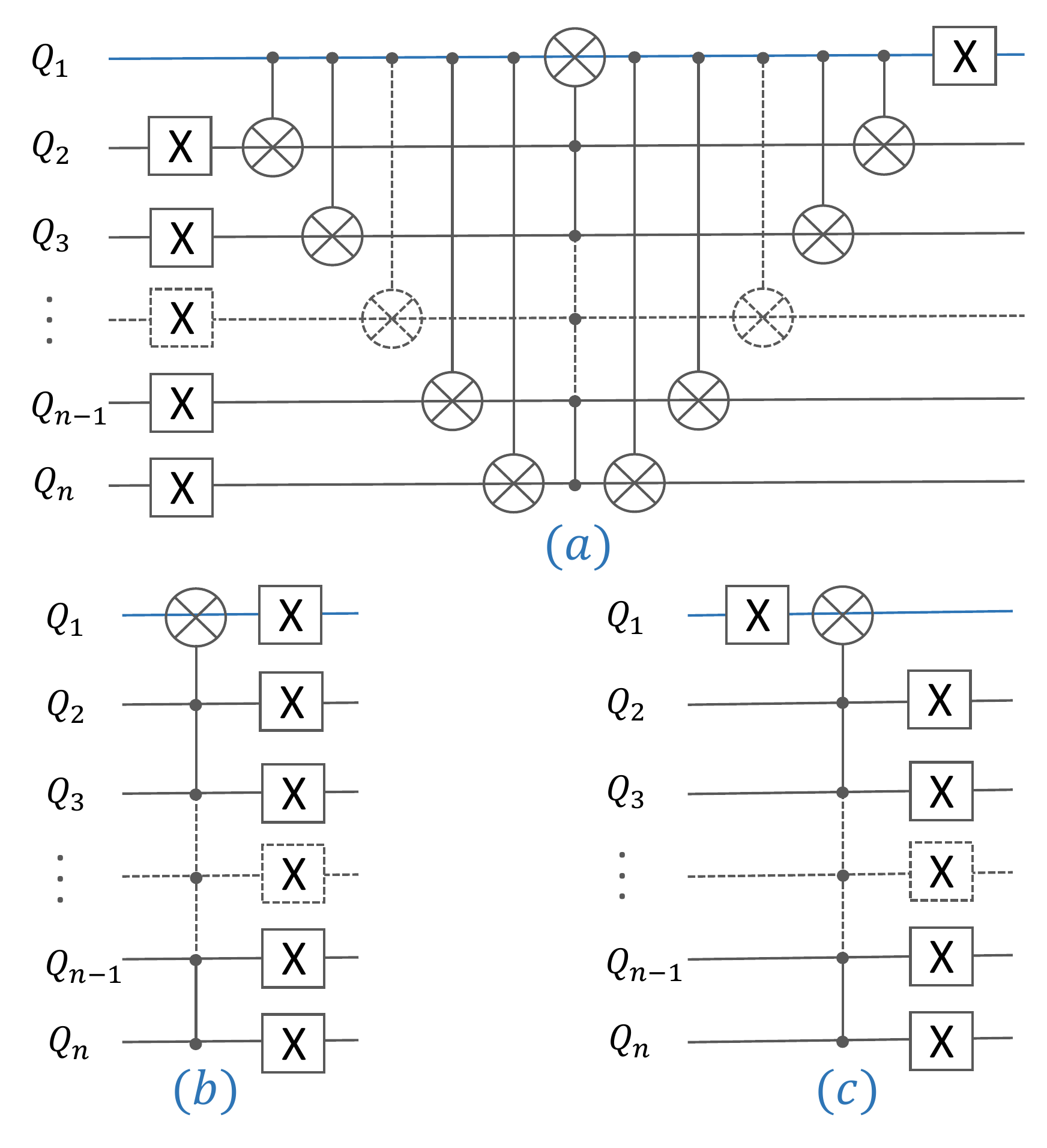}
\end{center}
\setlength{\abovecaptionskip}{-0.00cm}
\caption{\footnotesize{Quantum circuits for realizing the operations $U^m_n$. (a) Realizing $U^1_n$ in LPPS-TT1. (b) Realizing $U^2_n$ in LPPS-TT2. (c) Realizing $U^3_n$ in LPPS-TT3. The blue and black lines respectively represent the ancilla spin $Q_1$ and the work spins $Q_i$ $(i=2,...,n)$, and X denotes the $\sigma_x$ operation for flipping the spin. The controlled-not gate and Toffoli gate are performed conditional on the control spins being in the state $\ket{1}$ represented by gray circles. }}\label{circuit_detail}
\end{figure}

\textbf{LPPS-TT1}. The corresponding transformation $U^1_{n}$ is sparse zero-one matrix to redistribute the population, whose matrix can be found in Ref. \cite{sum}. For instance, for a 2-spin system, the thermal equilibrium state $\rho^1_2=\frac{1}{2}$Diag$(\gamma_1+\gamma_2,\gamma_1-\gamma_2,-\gamma_1+\gamma_2,-\gamma_1-\gamma_2)$ will be redistributed by $U^1_{2}$ to the density matrix $\rho^1_{2u}=\frac{1}{2}$Diag$(\gamma_1+\gamma_2,-\gamma_1+\gamma_2,\gamma_1-\gamma_2,-\gamma_1-\gamma_2)$. Combining with $\rho^1_2$ will create the desired PPS $\bar{\rho}^1_2=\rho^1_2+\rho^1_{2u}=(\gamma_1+\gamma_2)$Diag$(1,0,0,-1)$. In general, The method LPPS-TT1 steadily provides the experimentalist with a high-quality PPS $(\sum_{i}^{n}\gamma_i)(\ket{0}\bra{0}^{\otimes n}-\ket{1}\bra{1}^{\otimes n})$ via the simple two-step operations for any molecules. Additionally, This method actually produces two available LPPS, a PPS $\ket{0}\bra{0}^{\otimes n-1}$ in the subspace labeled by the state $\ket{0}$ of the ancilla qubit, and a state $\ket{1}\bra{1}^{\otimes n-1}$ labeled by the state $\ket{1}$. As shown in Fig. \ref{circuit_detail}, we present a quantum circuit for realizing the transforming $U^1_{n}$ in quantum computing networks by decomposing $U^1_{n}$ into $n$ single-qubit rotations $X=\sigma_x$, $2(n-1)$ controlled not gates, and a $n$-qubit Toffoli gate. These gates can be further realized by  single-qubit rotations and $J$-coupling evolutions in NMR techniques. It is undoubtedly worth noting that the increasing factor $\sum_{i}^{n}\gamma_i$ of the signal is obtained at the cost of an ancilla qubit and more quantum gate operations.

\textbf{LPPS-TT2(TT3)}. In some cases, it is more reasonable to prepare the following PPS experimentally, upto a factor, $(\ket{0}\bra{0}-\ket{1}\bra{1})\otimes \ket{0}\bra{0}^{\otimes (n-1)}$,  where the $\ket{0}$ and $\ket{1}$ states of the ancilla spin both label the  state $\ket{0}\bra{0}^{\otimes (n-1)}$. The reason lies in the fact that the signal of LPPS after a single-spin selective $\pi/2$ pulse on the ancilla spin is more identifiable in a NMR spectrum, and this labeling relationship can be actually exploited as a double-check for quantum computing. We provide two approaches LPPS-TT2 and LPPS-TT3 to realize this purpose.

Considering a 2-spin system as an example, $U^2_{2}$ transfers the density matrix $\rho^2_2=\gamma_1 I_z ^{1}+\gamma_2 I_z ^{2}$ to $\rho^2_{2u}=\frac{1}{2}$Diag$(\gamma_1-\gamma_2,-\gamma_1+\gamma_2,-\gamma_1-\gamma_2,\gamma_1+\gamma_2)$, leading to the desired PPS $\bar{\rho}^2_2=\rho^2_2+\rho^2_{2u}=\gamma_1$Diag$(1,0,-1,0)$. While, $\rho^3_2=\gamma_1 I_z ^{1}$ is changed  to $\rho^2_{2u}=\frac{1}{2}$Diag$(\gamma_1,-\gamma_1,-\gamma_1,\gamma_1)$ under the operation $U^3_{2}$, finally, the PPS $\bar{\rho}^3_2=\rho^3_2+\rho^3_{2u}=\gamma_1$Diag$(1,0,-1,0)$ is obtained. In general cases,  all of there methods successfully prepare a PPS $\bar{\rho}^2_n (\bar{\rho}^3_n) =\gamma_1(\ket{0}\bra{0}-\ket{1}\bra{1})\otimes \ket{0}\bra{0}^{\otimes (n-1)}$.  The general matrices of the reconstructed $U^2_{n}$ and $U^3_{n}$ can be found in Ref. \cite{sum}.
Fig. \ref{circuit_detail}(b) and Fig. \ref{circuit_detail}(c) show available quantum circuits for realizing $U^2_{n}$ and $U^3_{n}$, respectively. It is found that quantum circuit of $U^3_{n}$ is more complicated than that of $U^2_{n}$ when the same purpose is achieved. However, it appears that LPPS-TT3 gives more power into quantum computing in preparing the PPS, and gradient field in LPPS-TT3 can effectively destroy the undesired magnetization in $x$-$y$ plane.

\textit{Experiments. }-- Experimentally, we consider a heteronuclear 2-spin and a homonuclear 4-spin samples in liquid NMR, in order to illustrate the basic ideas of the new methods LPPS-TT$m$ independent of the structure of the molecules. We focus on the traceless matrices, the deviation density matrices, in the whole experiments. 

\textbf{Hetero-nuclear 2-spin case}.  The physical system to demonstrate the above processings was $^{13}$C-labeled chloroform \cite{tao1}. Nuclear spins of $^{13}$C and $^{1}$H encode the ancillary qubit and the work qubit, respectively. The corresponding parameters of the measured Hamiltonian $\mathcal{H}_2$ can be found in Ref. \cite{sum}, such as, the chemical shifts $\nu_i$ and the J-coupling constants $J_{ij}$.

\begin{figure*}[htb]
\begin{center}
\includegraphics[width= 1.6\columnwidth]{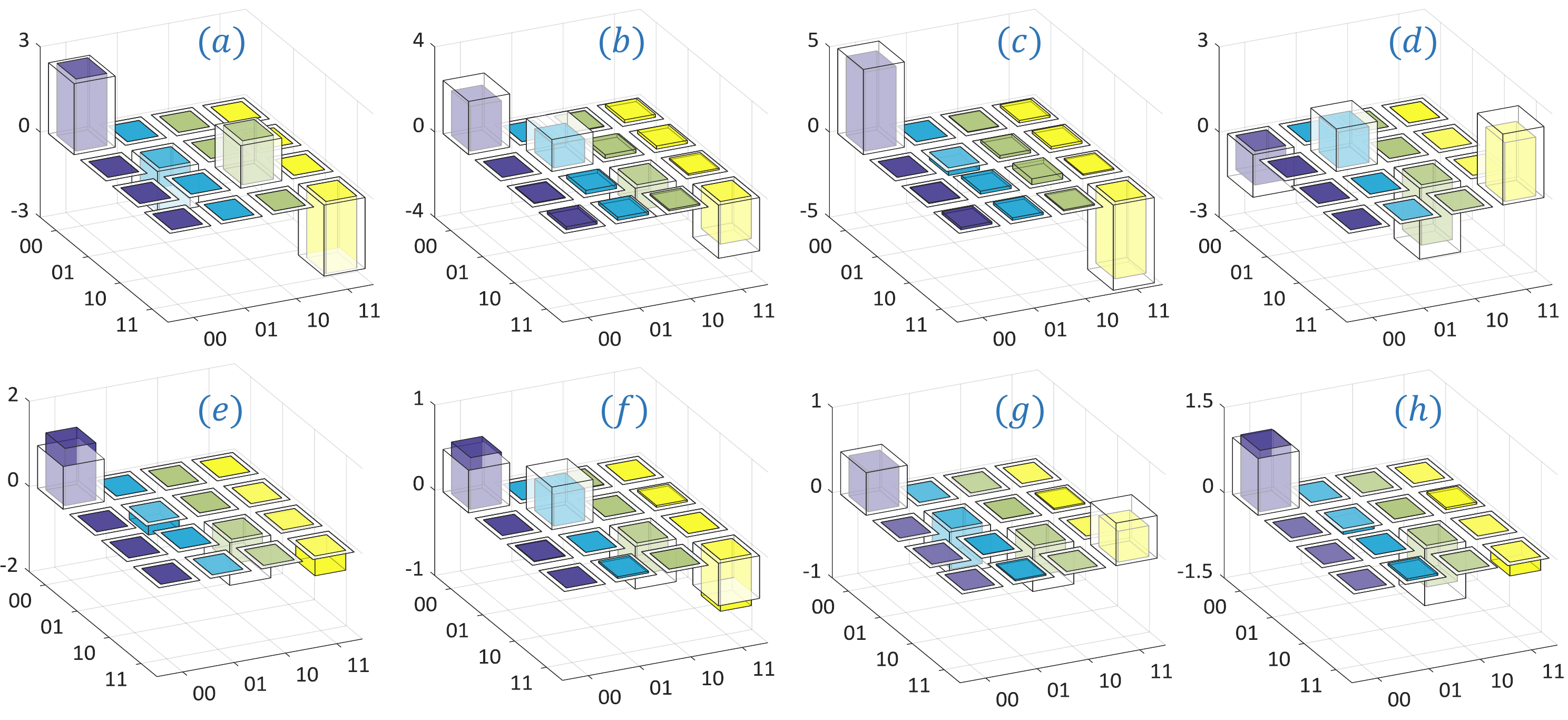}
\end{center}
\setlength{\abovecaptionskip}{-0.00cm}
\caption{\footnotesize{\textbf{Real parts of the reconstructed density matrices by full state tomography on $\rho^m_2$ and $\rho^m_{2u}$ (assuming that $\gamma_2=4\gamma_1=4$).} (a)-(c) respectively show the results of the reconstruction on  $\rho^1_2$, $\rho^1_{2u}$, and $\rho^1_2+\rho^1_{2u}$. (d) and (e) separately present the reconstructed density matrix $\rho^2_{2u}$ and $\rho^1_2+\rho^2_{2u}$. (f)-(g) illustrate the reconstructed results via the full state tomography on  $\rho^3_2$, $\rho^3_{2u}$, and $\rho^3_2+\rho^3_{2u}$, respectively. All the white bars are the theoretical results and the corresponding results are illustrated by the solid bars. }}\label{TT123_CH}
\end{figure*}

\begin{figure}[htb]
\begin{center}
\includegraphics[width= 0.7\columnwidth]{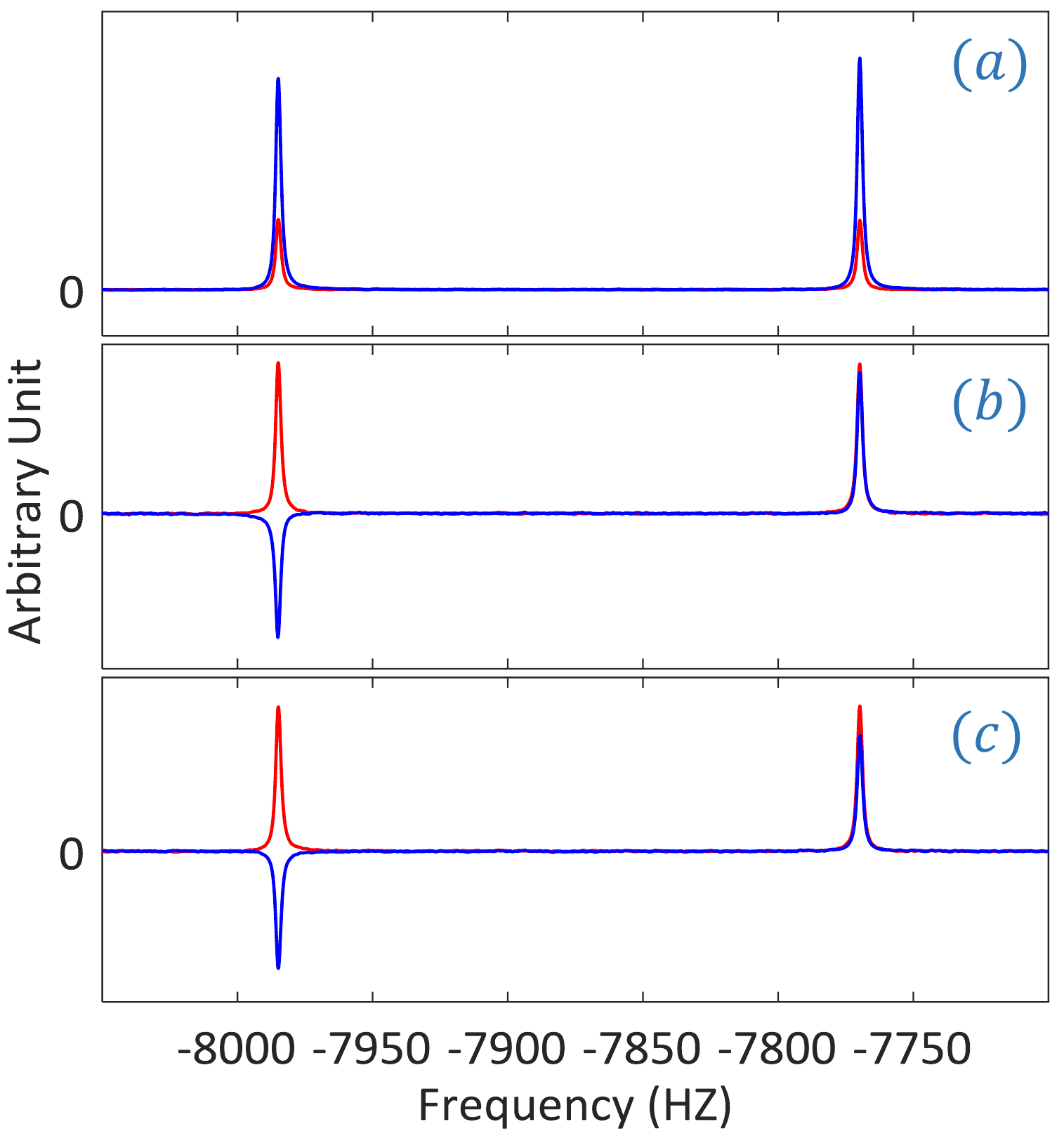}
\end{center}
\setlength{\abovecaptionskip}{-0.00cm}
\caption{\footnotesize{\textbf{Experimental spectra of the spin $^{13}$C (arbitrary units) for (a) LPPS-TT1, (b) LPPS-TT2, and (c) LPPS-TT3. } The red line in each subfigure represents the result after the single-qubit rotation $[\pi/2]^1_y$ pulse is applied on the final density matrix $\rho^m_2$, while the black lines mean the spectra by using $[\pi/2]^1_y$ pulse on the spin $^{13}$C behind the implementation of $U^m_2$ on the density matrix $\rho^m_2$. }}\label{CH_spec}
\end{figure}
First, the natural thermal state $\gamma_1 I_z ^{1}+\gamma_2 I_z ^{2}$ can be straightly used as the input as state $\rho^m_2$ for LPPS-TT$m$ $(m=1,2)$.  Applying a $\pi/2$ pulse on the spin $^{1}$H followed by a gradient field creates the input state $\rho^3_2=\gamma_1 I_z ^{1}$ for LPPS-TT3. Second, the implementation of  operations $U^m_2$ are decomposed into the following pulse sequences,
\begin{eqnarray}
&& U^1_2 : [\pi]^2_x \rightarrow [\text{SWAP}] \rightarrow [\pi]^1_x, \nonumber \\
&& [\text{SWAP}] : [-\pi/2]^{1,2}_x \rightarrow [1/2J] \rightarrow [\pi/2]^{1,2}_x \rightarrow [-\pi/2]^{1,2}_y \nonumber \\
&& \rightarrow [1/2J]  \rightarrow [\pi/2]^{1,2}_y , \nonumber \\
&& U^2_2 : [\pi/2]^1_y \rightarrow [1/2J] \rightarrow [-\pi/2]^1_x \rightarrow [\pi]^2_x,\nonumber \\
&& U^3_2 : [\pi]^1_x \rightarrow [\pi/2]^1_y \rightarrow [1/2J] \rightarrow [\pi/2]^1_x \rightarrow [\pi]^2_x, 
\label{decompose}
\end{eqnarray}
where $[\theta]^k_\alpha$ means a $\theta$  rotation around $\alpha$
direction on the spin $k$, and $[1/2J]$ represents the free evolution $e^{-i\pi I^1_zI^2_z}$. In principle, the  controlled-not gates $U_{a,b}$ in Fig. \ref{circuit_detail}, where qubit $a$ and $b$ respectively mean the control and target qubit,  should be decomposed into the sum of the local rotations and the J-coupling evolution,
\begin{eqnarray}
U_{a,b} : [\frac{\pi}{2}]^b_y \rightarrow [1/2J] \rightarrow [\frac{\pi}{2}]^b_x \rightarrow [-\frac{\pi}{2}]^b_z \rightarrow [\frac{\pi}{2}]^a_z. 
\label{decompose1}
\end{eqnarray}
We optimized the whole pulse sequence to obtain simplified pulses illustrated in equation (\ref{decompose}). The reason why it is  feasible is based on the fact that some operations do not have any influence on some traceless elements of density matrix theoretically, such as operations $[\pi]^1_x$  acting on the element $I_xI$. Besides, the simplified sequence usually is more robust again the imprecision of the operations than the virgin sequences.

Experimental spectra of the demonstration of the methods LPPS-TT$m$ were obtained, and Fig. \ref{CH_spec} shows $^{13}$C spectra by applying a spin-selective $[\pi]^1_y$ pulse on the final density matrix $\rho^m_2$ and $\rho^m_{2u}$. The obtained spectra clearly show that the LPPS-TT1 method has a strong ability to increase the signal-noise ratio, and only one peak is visible in the methods LPPS-TT2 and LPPS-TT3. Besides, we performed two-qubit full state tomography on the states  $\rho^m_{2}$ and $\rho^m_{2u}$ \cite{Leskowitz,Lee}. The real part of the reconstructed density matrices are illustrated in Fig. \ref{TT123_CH}, which definitely affirms that the logical labeled PPS $(\ket{00}\bra{00}-\ket{11}\bra{11})$ and $(\ket{00}\bra{00}-\ket{10}\bra{10})$ are successfully prepared via the proposed LPPS-TT$m$ methods. It is worth emphasizing that we merely reconstruct the thermal state once as the results of $\rho^1_2$ and $\rho^2_2$ because LPPS-TT1 and LPPS-TT2 are both based on the thermal state.  
\begin{figure}[htb]
\begin{center}
\includegraphics[width= 0.8\columnwidth]{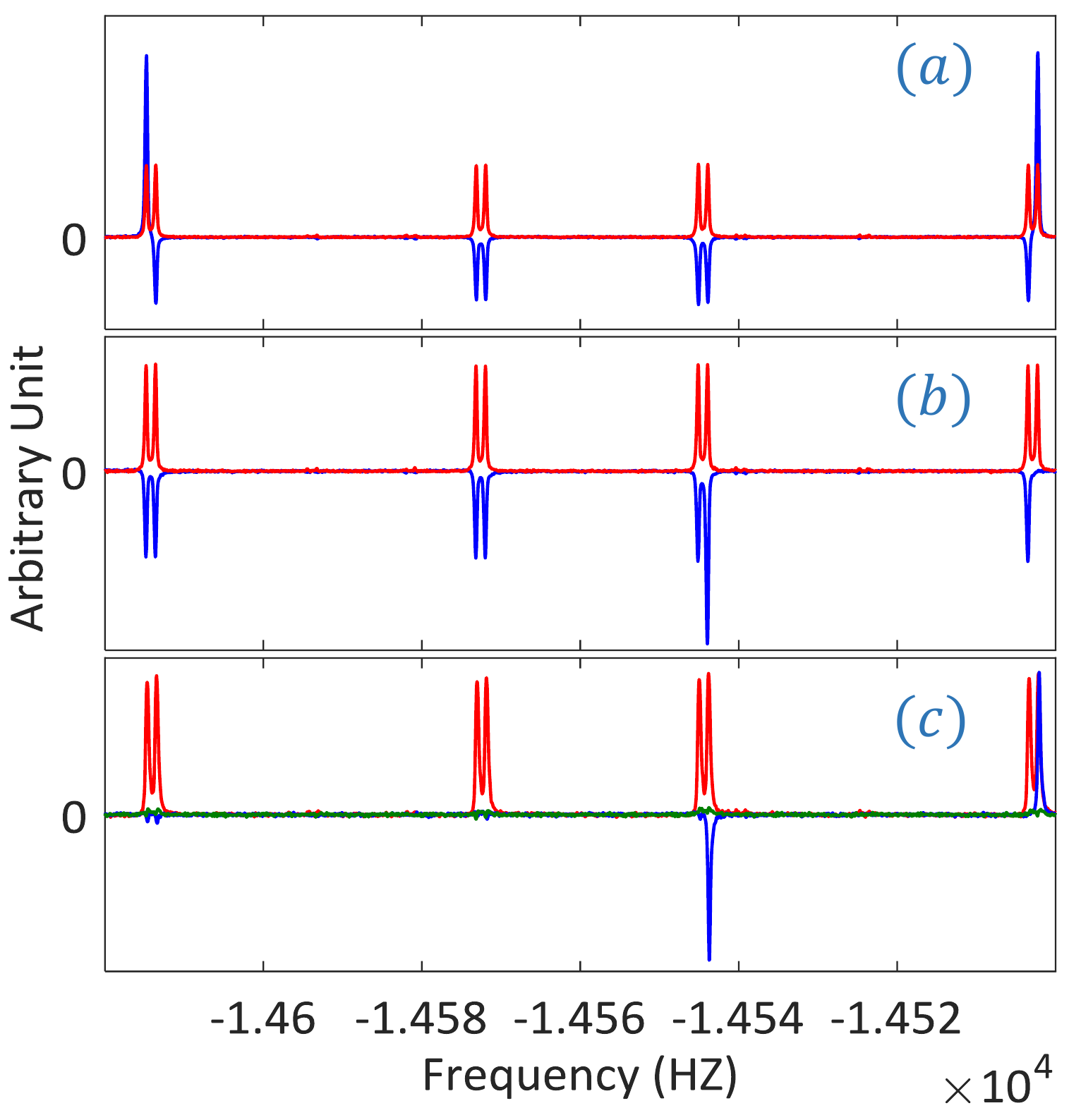}
\end{center}
\setlength{\abovecaptionskip}{-0.00cm}
\caption{\footnotesize{\textbf{Experimental spectra of the spin $^{13}$C2 (arbitrary units) for (a) LPPS-TT1, (b) LPPS-TT2, and (c) LPPS-TT3. } The red line in each subfigure shows the thermal spectrum of nuclear $^{13}$C2 in which the single-qubit rotation $[\pi/2]^2_y$ pulse is applied on thermal density matrix $\sum^4_{i=1}\gamma_i I_z ^{1}$, while the black lines mean the spectra by using $[\pi/2]^2_y$ pulse on the spin $^{13}$C2 behind the implementation of $U^m_4$ on the density matrix $\rho^m_{4u}$.  The cyan line of Plot (c) is observed by applying $[\pi/2]^2_y$ pulse on the input state of LPPS-TT3  $\rho^3_4=\gamma_1 I_z ^{1}$. }}\label{4C_spec}
\end{figure}

\textbf{Homo-nuclear 4-spin case}. In order to demonstrate our proposal being independent of the structure of the used molecules, we consider a homo-nuclear 4-spin system to demonstrate the methods LPPS-TT$m$, which is $^{13}$C-labeled trans-crotonic acid dissolved in d6-acetone \cite{tao2}. The structure of the molecule is shown in Ref. \cite{sum}. C$_1$ to C$_4$ correspondingly denote the four qubits Q$_1$ to Q$_4$. We decoupled the methyl group M, H$_1$ and H$_2$ throughout all experiments.

\begin{figure}[htb]
\begin{center}
\includegraphics[width= 0.8\columnwidth]{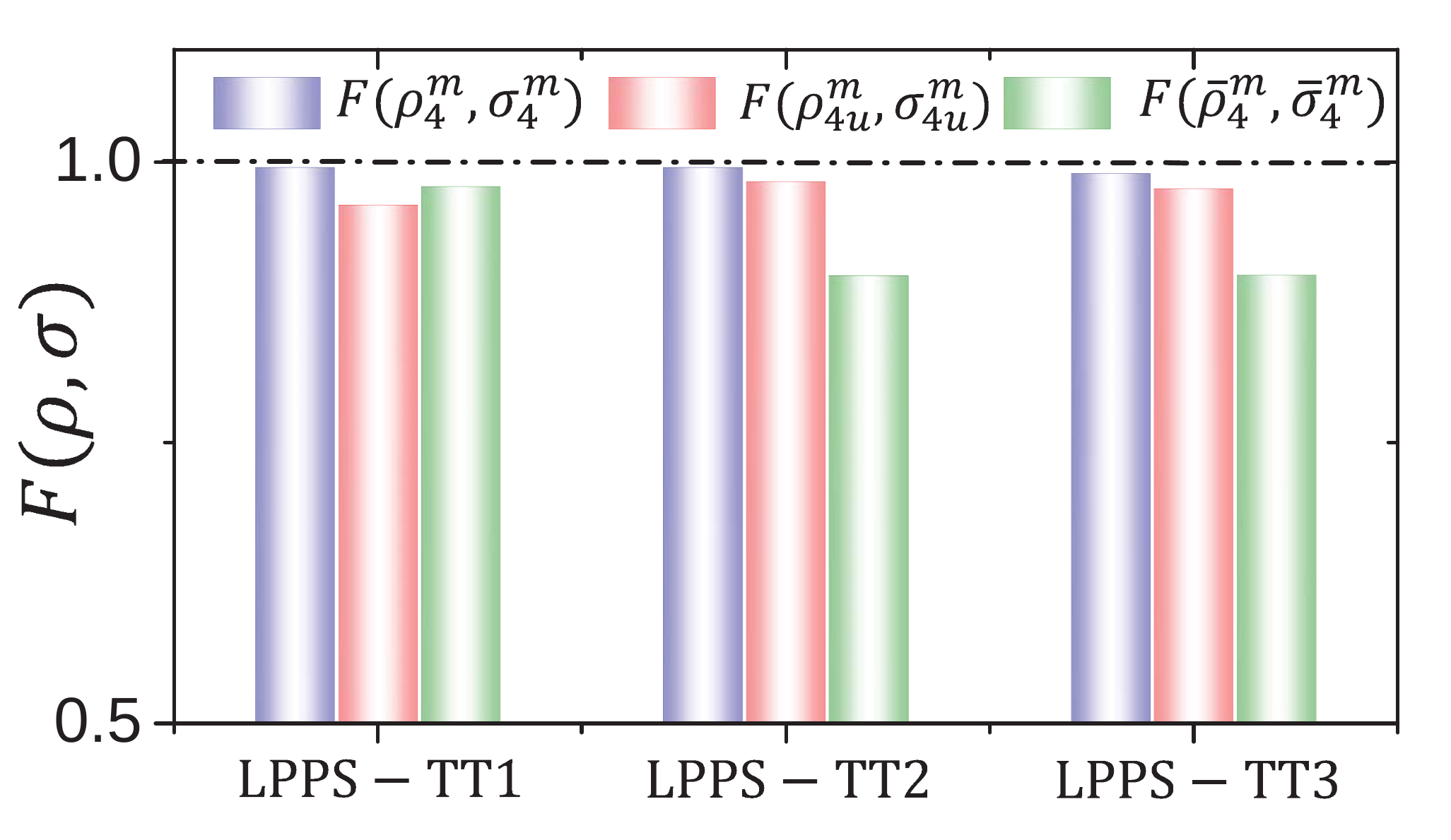}
\end{center}
\setlength{\abovecaptionskip}{-0.00cm}
\caption{\footnotesize{\textbf{Comparison of the different methods for preparing the PPS. }  The blue, red and cyan bars respectively show the fidelities $F(\rho^m_4,\sigma^m_4)$, $F(\rho^m_{4u},\sigma^m_{4u})$, and $F(\bar{\rho}^m_4 ,\bar{\sigma}^m_4)$, which was calculated by the definition $F(\rho,\sigma)=|\text{Tr}(\rho \sigma)|/\sqrt{\text{Tr}(\rho^2)\text{Tr}(\sigma^2)}$.  }}\label{4C_F}
\end{figure}

Similar to Hetero-nuclear 2-spin case, the natural thermal state $\sum^4_{i=1}\gamma_i I_z ^{i}$ was  used as the input  $\rho^m_4$ for LPPS-TT$m$ $(m=1,2)$, and $\gamma_1 I_z ^{1}$ was used as the input  $\rho^3_4$, which was also realized by a single-selective $\pi/2$ pulse on the spins C2 to C4 followed by a gradient field. Instead of using the decomposed pulse sequence, we realize the desired operations $U^m_4$ via GRadient Ascent Pulse Engineering (GRAPE) techniques which provides a 50ms shaped-pulse width and over 99.5\% fidelity \cite{Khaneja,Ryan}, because too many pulses usually lead to accumulation of  the imprecision if the quantum circuits for realizing $U^m_4$ are decomposed into single-qubit rotations and the $J$-coupling evolutions. For each density matrix $\rho^m_4$ and $\rho^m_{4u}$, we apply a spin-selective $[\pi/2]^2_y$ pulse on the spin C2 to obtain the spectra shown in Fig. \ref{4C_spec}. The corresponding spectra of the desired $\bar{\rho}^m_4$ can be directly obtained by summing over the spectra of $\rho^m_4$ and $\rho^m_{4u}$, in which merely two peaks respectively labeled by the $\ket{1}$ and $\ket{0}$ of the ancilla spin from right to left were observed. Meanwhile, we further performed 4-qubit full state tomography on the density matrices $\rho^m_4$ and $\rho^m_{4u}$ \cite{Leskowitz}, and reconstructed them as $\sigma^m_4$ and $\sigma^m_{4u}$, with $\bar{\sigma}^m_{4}=\sigma^m_{4}+\sigma^m_{4u}$. Fig. \ref{4C_F} illustrates the distance between $\rho$ and $\sigma$ for all methods LPPS-TT$m$, in which the averaging fidelity is over 96.4\%.

\textit{Conclusion. }--
 In this work, we propose  an efficient scheme LPPS-TT$m$ for preparing pseudo-pure state for bulk quantum computing. It combines the advantage of logical
labeling and temporal averaging method. Compared with existing schemes of initialization which use either exponential number of turns of execution, or cause signal reduction or places restriction on the molecular structure, Our proposed methods merely use two turns of execution, irrespective the number of qubits, and have no restrictions on the
molecular structure of the samples, which have greatly simplify the initialization procedure for bulk quantum computing with large number of qubits. The method LPPS-TT1 has a stronger signal, because the factor $\sum_i\gamma_i$ are created compared with the method LPPS-TT2(TT3) where only $\gamma_1$ has a contribution for the signal. However, LPPS-TT2(TT3) has the benefit of providing a simple spectrum structure which is convenient for some quantum algorithms such as the Grover algorithm. Experimentally, we consider a heteronuclear 2-spin and a homonuclear 4-spin as examples to demonstrate the processing of preparing logical PPS via our methods. The results surely show that the desired PPS is successfully prepared with the high quality. The scheme is a general scheme and is not restricted to bulk quantum computation such as a liquid NMR. They may also be applied to other quantum computer platforms to eliminate the effects of noises by increasing the signal-noise ratio.

\begin{acknowledgments}
{\bf Acknowledgments.} T. X. and G. L. are grateful to the following funding sources: National Natural Science Foundation of China under Grants No. 11175094 and No. 91221205; National Basic Research Program of China under Grant No. 2015CB921002. S. Y. H is supported by the Science Challenge Project (SCP) under Grant No. TZ2016003-1.
\end{acknowledgments}

\clearpage

\onecolumngrid

\section*{Supplemental Material for \\ ``Preparation of Logically Labeled Pure States with Only Two Turns for Bulk Quantum Computation''}

Further experimental details and results, as well as the matrix form of the operation $U^m_n$, are provided in this Supplemental Material. 
\\

{\it{\bfseries{The reconstruction of $U^m_n$}}}-- In this work, we proposed a novel framework to prepare logically labeled pure states with only two turns for bulk quantum computation. The unitary operation $U^m_n$ is reconstructed to redistribute the population of $\rho^m_n$ in second step, such that the desired logically labeled PPS $\bar{\rho}^m_n=\rho^m_n+\rho^m_{nu}$ is obtained by combining the input state $\rho^m_n$ in first step. It is worth emphasizing that $U^m_n$ is a sparse and structured zero-one matrix. The general forms of the reconstructed $U^m_{n}$ can be described as, 

\begin{equation}\label{U1}
U^1_{n}=\left[\begin{array}{ccccccc} 1 &   &   &   &   &   &   \\  &   &   &   &   & 1  &   \\  &   &   &   & \cdot  &   &   \\  &   &   &  \cdot &   &   &   \\  &   &  \cdot &   &   &   &   \\  & 1  &   &   &   &   &   \\  &   &   &   &   &   & 1 \end{array}\right], 
U^2_{n}=\left[\begin{array}{cccccc}  &   &  1 &   &   &   \\  &   &   &   & \cdot  &   \\  &   &   &  1 &   &   \\  &   &   &   &   &  1 \\  &  \cdot &   &   &   &   \\  1&   &   &   &   &  \end{array}\right],
U^3_{n}=\left[\begin{array}{cccccc} 1 &   &   &   &   &   \\  &   &   &   & \cdot  &   \\  &   &   &   &   &  1 \\  &   &   & 1  &   &  \\  &  \cdot &   &   &   &   \\  &   &  1 &   &   &  \end{array}\right].
\end{equation}
where the empty entries are all filled with zeroes. For a 4-spin system, the matrices of $U^m_{4}$ are,
\begin{equation}\label{Uu}
U^1_{4}=\left[\begin{array}{cccccccc}1 & 0 & 0 & 0 & 0 & 0 & 0 & 0 \\0 & 0 & 0 & 0 & 0 & 0 & 1 & 0 \\0 & 0 & 0 & 0 & 0 & 1 & 0 & 0 \\0 & 0 & 0 & 0 & 1 & 0 & 0 & 0 \\0 & 0 & 0 & 1 & 0 & 0 & 0 & 0 \\0 & 0 & 1 & 0 & 0 & 0 & 0 & 0 \\0 & 1 & 0 & 0 & 0 & 0 & 0 & 0 \\0 & 0 & 0 & 0 & 0 & 0 & 0 & 1\end{array}\right],
U^2_{4}=\left[\begin{array}{cccccccc}0 & 0 & 0 & 1& 0 & 0 & 0 & 0 \\0 & 0 & 0 & 0 & 0 & 0 & 1 & 0 \\0 & 0 & 0 & 0 & 0 & 1 & 0 & 0 \\0 & 0 & 0 & 0 & 1 & 0 & 0 & 0 \\0 & 0 & 0 & 0 & 0 & 0 & 0 & 1 \\0 & 0 & 1 & 0 & 0 & 0 & 0 & 0 \\0 & 1 & 0 & 0 & 0 & 0 & 0 & 0 \\1 & 0 & 0 & 0 & 0 & 0 & 0 & 0\end{array}\right],
U^3_{4}=\left[\begin{array}{cccccccc}1 & 0 & 0 & 0 & 0 & 0 & 0 & 0 \\0 & 0 & 0 & 0 & 0 & 1 & 0 & 0 \\0 & 0 & 0 & 0 & 0 & 0 & 1 & 0 \\0 & 0 & 0 & 0 & 0 & 0 & 0 & 1 \\0 & 0 & 0 & 0 & 1 & 0 & 0 & 0 \\0 & 1 & 0 & 0 & 0 & 0 & 0 & 0 \\0 & 0 & 1 & 0 & 0 & 0 & 0 & 0 \\0 & 0 & 0 & 1 & 0 & 0 & 0 & 0\end{array}\right]
\end{equation}
\\
\begin{figure}[htb]
\begin{center}
\includegraphics[width= 0.5\columnwidth]{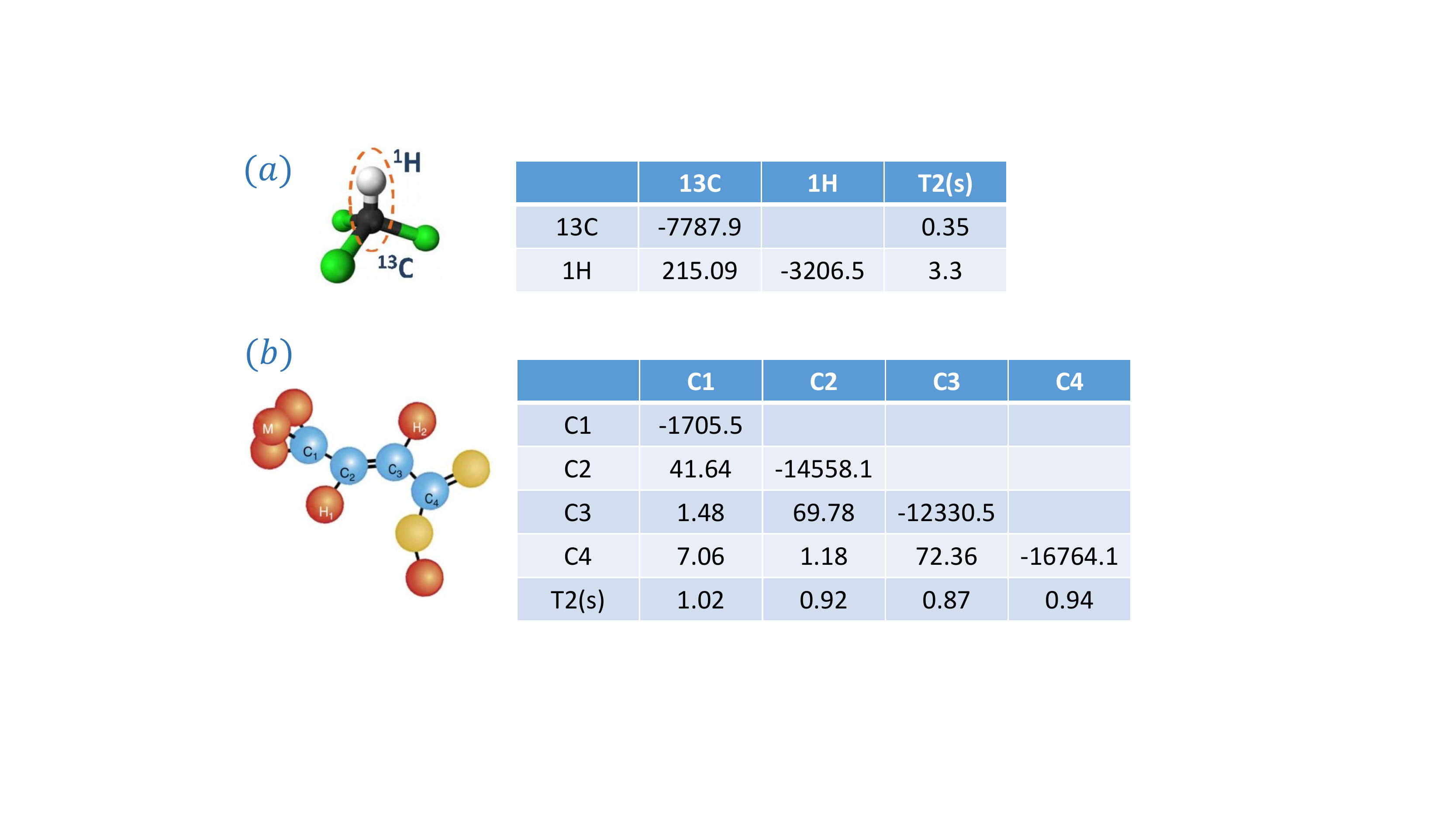}
\end{center}
\setlength{\abovecaptionskip}{-0.00cm}
\caption{\footnotesize{\textbf{Molecular structure and relevant parameters of $^{13}$C-labeled Chloroform (a) and $^{13}$C-labeled trans-crotonic acid (b).} Diagonal elements and off-diagonal elements in the table provide the values of the chemical shifts (Hz) and $J$-coupling constant (Hz) between different nuclei of the molecule. The table also provides transversal relaxation $T_2$ , which can be measured using the standard inversion recovery and Hahn echo sequences. }}\label{molecule}
\end{figure}

{\it{\bfseries{Experimental samples}}} --In experiments, we have employed the  sample of $^{13}$C-labeled chloroform dissolved in d6-acetone as hetero-nuclear 2-spin case. Analogously, $^{13}$C-labeled trans-crotonic acid dissolved in d6-acetone is used as a homo-nuclear 4-spin case, as indicated in the main text. 
In figure~\ref{molecule} we give a pictorial representation of the molecule structure together with the values of some relevant parameters, such as, the chemical shifts $\nu_i$ and the J-coupling constants $J_{ij}$.
\\

\begin{figure}[htb]
\begin{center}
\includegraphics[width= 0.8\columnwidth]{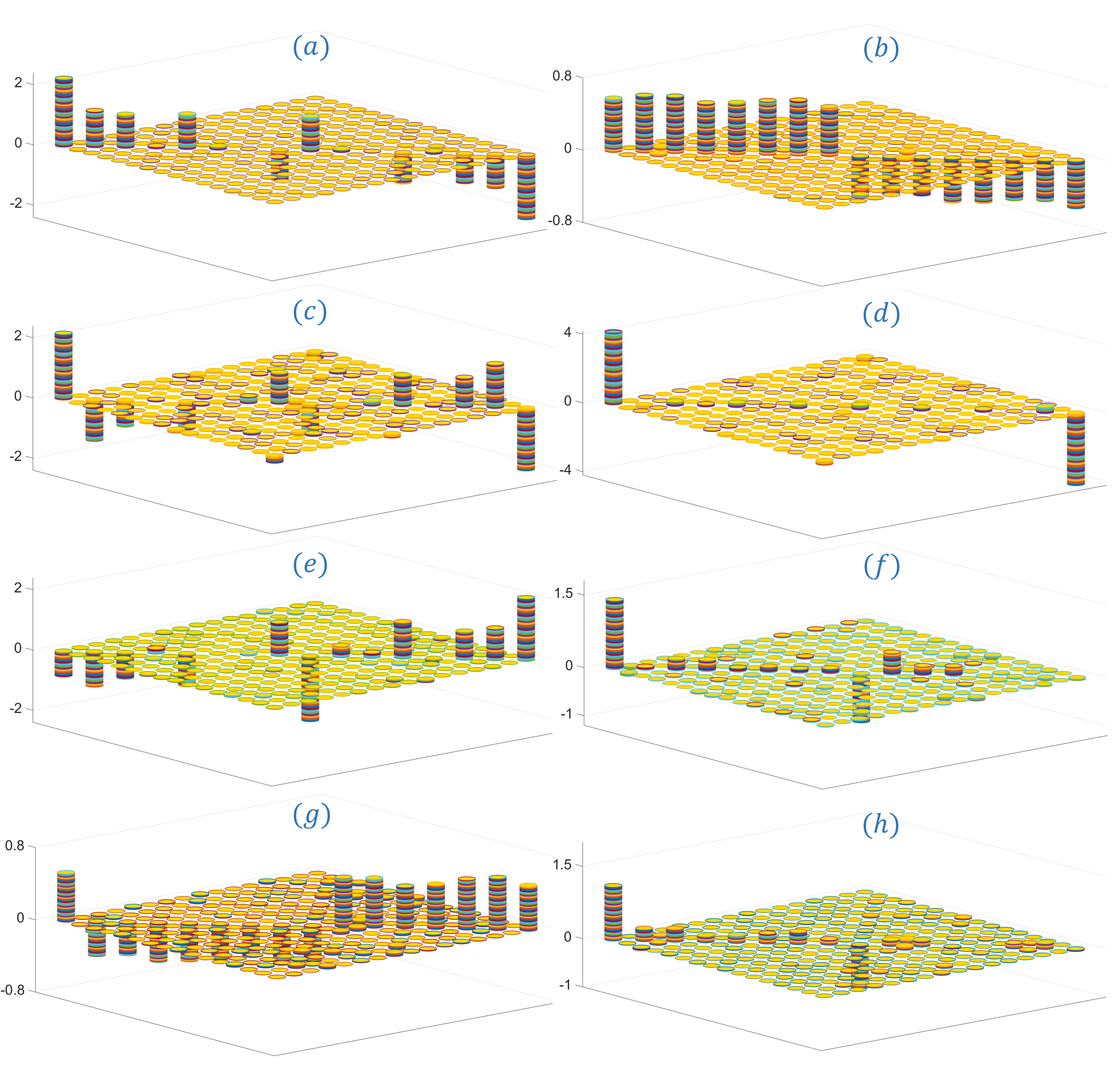}
\end{center}
\setlength{\abovecaptionskip}{-0.00cm}
\caption{\footnotesize{\textbf{Real parts of the reconstructed density matrices $\sigma^m_4$ and $\sigma^m_{4u}$ (assuming that $\gamma_i=1$).}  Plots (a) and (b) respectively show the reconstructed results of density matrices $\sum^4_{i=1}\gamma_i I_z ^{i}$ and $\gamma_1 I_z ^{1}$. (c) and (d) present the real parts of the reconstructed density matrices $\sigma^1_{4u}$ and $\bar{\sigma}^1_4$ with $\bar{\sigma}^1_4=\sigma^1_{4}+\sigma^1_{4u}$ for the method LPPS-TT1. Analogously, (e) and (f) show the reconstructed results $\sigma^2_{4u}$ and $\bar{\sigma}^2_4$ for  LPPS-TT2. (g) and (h) show the reconstructed results $\sigma^3_{4u}$ and $\bar{\sigma}^3_4$ for  LPPS-TT3. }}\label{matrix4}
\end{figure}

{\it{\bfseries{Full state tomography of $\rho^m_4$ and $\rho^m_{4u}$}}} --In the main text, we performed 4-qubit full state tomography on the density matrices $\rho^m_4$ and $\rho^m_{4u}$ and reconstructed them as $\sigma^m_4$ and $\sigma^m_{4u}$ via the following observable pulses,
\begin{eqnarray}
&& \text{Pulse~set} : [XXXX,IIYY,YYXX,IIIY,XYXX,YXYI,IXYI, \nonumber \\
&& IIIX,XIYY,YXII,YYXY,XYXI,IIYX,IXIY,IIXI,IYIY]
\label{decompose pulse}
\end{eqnarray}
Here, $X=\exp(-i\sigma_x\pi/4)$, $Y=\exp(-i\sigma_y\pi/4)$ and $I$ is a $2\times 2$ identity operation. For instance, an observable pulse $IXYI$ means the operation $I\otimes \exp(-i\sigma^2_x\pi/4) \otimes \exp(-i\sigma^3_y\pi/4) \otimes I$ is applied on the reconstructed density matrix. Figure ~\ref{matrix4} shows the real parts of the reconstructed density matrices $\sigma^m_4$, $\sigma^m_{4u}$ and $\bar{\sigma}^m_{4}$.

\end{document}